\documentclass[aps,onecolumn,nofootinbib,superscriptaddress,letterpaper,amsmath,amssymb]{revtex4}
\pdfoutput=1

\usepackage{graphicx}  
\usepackage{dcolumn}   
\usepackage{bbm}        
\usepackage{amsmath}
\usepackage{amsfonts}
\usepackage{amssymb}   
\usepackage{epstopdf}
\usepackage{slashed}
\usepackage{hyperref}

\usepackage{multirow}
\usepackage{color}
\usepackage{float}
\usepackage{ragged2e}
\usepackage{subcaption}
\usepackage{lineno}
\DeclareCaptionJustification{justified}{\justifying}
\usepackage[normalem]{ulem}
\usepackage{verbatim}
\usepackage{xspace}

\def\gsim{\lower0.5ex\hbox{$\:\buildrel >\over\sim\:$}}
\def\lsim{\lower0.5ex\hbox{$\:\buildrel <\over\sim\:$}}

\newcommand{\rev}[1]{#1}
\newcommand{\secref}[1]{{Sec.~\ref{#1}}}

\newcommand{\met}{$p_{\rm T}^{\rm miss}$\xspace}
\newcommand{\be}{\begin{equation}}
\newcommand{\ee}{\end{equation}}
\newcommand{\bea}{\begin{eqnarray}}
\newcommand{\eea}{\end{eqnarray}}

\newcommand{\nbox}{{\,\lower0.9pt\vbox{\hrule \hbox{\vrule height 0.2 cm
\hskip 0.2 cm \vrule height 0.2 cm}\hrule}\,}}

\newskip\zatskip \zatskip=0pt plus0pt minus0pt
\def\matth{\mathsurround=0pt}
\def\lsim{\mathrel{\mathpalette\atversim<}}
\def\gsim{\mathrel{\mathpalette\atversim>}}
\def\sigv{\ifmmode \langle\sigma v\rangle\else $\langle\sigma v\rangle$\fi}
\newskip\zatskip \zatskip=0pt plus0pt minus0pt
\def\matth{\mathsurround=0pt}
\def\lsim{\mathrel{\mathpalette\atversim<}}
\def\gsim{\mathrel{\mathpalette\atversim>}}
\def\atversim#1#2{\lower0.7ex\vbox{\baselineskip\zatskip\lineskip\zatskip
  \lineskiplimit
  0pt\ialign{$\matth#1\hfil##\hfil$\crcr#2\crcr\sim\crcr}}}

\begin{document}

\thispagestyle{empty}

\vspace{0.5in}

\title{Learning to Reconstruct Quirky Tracks}
\author{Qiyu Sha}
\affiliation{Institute of High Energy Physics, 19B Yuquan Road, Shijingshan District, Beijing 100049, China}
\affiliation{University of Chinese Academy of Sciences, 19A Yuquan Road, Shijingshan District, Beijing 100049, China}
\author{Daniel Murnane}
\affiliation{Niels Bohr Institute, University of Copenhagen, Denmark}
\affiliation{Scientific Data Division, Lawrence Berkeley National Laboratory, Berkeley CA}
\author{Max Fieg}
\affiliation{Department of Physics \& Astronomy, University of California, Irvine, CA}
\author{Shelley Tong}
\affiliation{Department of Physics \& Astronomy, University of California, Irvine, CA}
\author{Mark Zakharyan}
\affiliation{Department of Physics, University of California, Santa Barbara, CA}
\author{Yaquan Fang}
\affiliation{Institute of High Energy Physics, 19B Yuquan Road, Shijingshan District, Beijing 100049, China}
\affiliation{University of Chinese Academy of Sciences, 19A Yuquan Road, Shijingshan District, Beijing 100049, China}
\author{Daniel Whiteson}
\affiliation{Department of Physics \& Astronomy, University of California, Irvine, CA}

\begin{abstract}

Analysis of data from particle physics experiments traditionally sacrifices some sensitivity to new particles for the sake of practical computability, effectively ignoring some potentially striking signatures. However, recent advances in \rev{machine-learning} (ML)-based tracking allow for new inroads into  previously inaccessible territory, such as reconstruction of tracks \rev{that} do not follow helical trajectories. This paper presents a demonstration of the capacity of \rev{ML}-based tracking to reconstruct the oscillating trajectories of quirks, \rev{particles charged under an unbroken QCD-like gauge symmetry}. The technique used is not specific to quirks, and opens the door to a program of searching for many kinds of non-standard tracks.
\end{abstract}
\maketitle

\section{Introduction}
\label{Sec:Introduction}

Hadronic collisions at the energy frontier have wide power to reveal \rev{or constrain new physics~\cite{ATLAS:2012yve,CMS:2012qbp,annurev:/content/journals/10.1146/annurev-nucl-101920-014923,annurev:/content/journals/10.1146/annurev-nucl-102419-052854,LHCb:2015yax,LHCb:2014vgu,ATLAS:2010isq,CMS:2010ifv}}, such as new particles or interactions, whether expected or unexpected. In the coming years, the Large Hadron Collider (LHC) will accumulate a dataset of unprecedented luminosity, with an expected ${\cal L}=300 ~{\rm fb}^{-1}$ ($3 ~{\rm ab}^{-1}$) with the completion of Run 3 (Run 4+). While the current physics program is broad, the technical challenges of analyzing the high-dimensional, high volume dataset have historically required imposing assumptions on the nature of the potential new physics, which translate to strategic sacrifices in sensitivity, especially to the unexpected.   As a result, signatures of groundbreaking discoveries may be overlooked by the traditional analysis pipelines.

A crucial and challenging component of that pipeline is identifying tracks of charged particles in the inner, magnetized detector volume. Ionization leaves {\it hits} as particles pass through a tracking detector, but the task of assembling sets of hits to reconstruct {\it tracks} of particles is a challenging combinatorial problem. The task is made tractable, though still computationally costly, if one can assume that all tracks follow a helical trajectory and originate from within the beam pipe, as is expected of electrically charged particles created in an interaction at the beam crossing. This narrowed scope is sufficient for many expected signatures, but consequentially restricts the power to discover unexpected new physics.  This is well understood, and significant effort has been made to relax the assumption that tracks must originate inside the beam pipe by developing dedicated software to reconstruct tracks with displaced vertices \rev{(recent results include Refs.\cite{CMS:2024gxp, ATLAS:2023oti,CMS:2024xzb,Lee:2018pag,Alimena:2019zri,CMS:2024zqs})}, as might be expected from decays of long-lived particles.  Comparatively little effort~\cite{ATLAS:2023esy,CMS:2024eyx,CDF:2005cvf,Knapen:2017kly, CMS:2024eyx} has been put towards development of algorithms capable of identifying a potentially much broader class of tracks, those which are charged under other forces and do not follow helical paths in a solenoidal magnetic field. Such striking tracks may already exist in our data, but would be largely invisible to our current algorithms.

An important challenge in reconstructing such non-helical tracks is that traditional tracking algorithms~\cite{Strandlie:2010zz,ATLAS:2017kyn,CMS:2014pgm,Duda:1972ymn} assume the specific parametric form of the desired track in a way that does not trivially generalize to new forms. Each new potential trajectory would therefore require its own bespoke tracking algorithm  and incur its own computational expense in the analysis pipeline.   However, recent innovations in machine learning (ML) \rev{such as geometric deep learning~\cite{Bronstein:2016thv}} have allowed for the development of ML-based tracking algorithms which are architecturally agnostic to the parametric form of the track~\cite{ExaTrkX:2020nyf}, where the target class of track is defined only by the sample used to train the algorithm. In principle, this allows for rapid development of tracking pipelines which are more flexible, replacing the complex work of dedicated algorithm development with the much simpler task of preparation of a new training sample. If successful, it lays the groundwork for a more ambitious task, preparation of a generalized training sample that produces a track finder capable of reconstructing a broad class of helical and non-helical trajectories, with true sensitivity to unexpected tracks.

In this paper, we take the first step in that direction by demonstrating that ML-based tracking algorithms can be trained to identify non-helical tracks from a training sample, without the development of a dedicated new tracking algorithm. We train the \rev{graph-neural-network (GNN)} machine learning algorithm {\sc Exa.TrkX}~\cite{ExaTrkX:2021abe} to learn to reconstruct the strikingly non-helical oscillatory tracks produced by \textit{quirk pairs}. Quirks~\cite{Kang:2008ea} are  charged under an unbroken confining QCD-like gauge symmetry but, in contrast with quarks, the quirk masses are large relative to the confinement scale, resulting in an oscillatory non-helical motion by quirk pairs in a magnetic field. These spectacular signatures are such a departure from the classical helical tracks, given the in-flight interactions between the quirks, that even a bespoke application of the traditional algorithms has been a long-standing  challenge. Previous efforts have attempted to side-step this complexity by searching for straight tracks~\cite{D0:2010kkd}, momentum imbalance~\cite{Farina:2017cts}, stopped out-of-time particles~\cite{Evans:2018jmd} or sets of hits along a plane~\cite{Knapen:2017kly}, essentially ignoring the oscillatory nature of the tracks. This is reasonably effective for a limited quirk parameter range, but we show that ML-based tracking is more broadly efficient without requiring such quirk-specific simplifications.  Our studies also demonstrate a capacity to generalize beyond examples in the training set. This gives hope that a similar approach could be used to  construct a pipeline that can reconstruct tracks from a broader class of particles, unrestricted to helical tracks or quirks. This  may pave the way for discoveries at the LHC that have evaded our detection. Complementary studies show sensitivity to quirks in the far-forward region with experiments such as FASER~\cite{Feng:2024zgp,Li:2021tsy,Li:2023jrt,MammenAbraham:2024gun}.

The paper is organized as follows. In \secref{Sec:Background} we cover the background of our study by reviewing the quirk model in a collider setting. In \secref{Sec:Modeling}, we discuss the equations of motion for the quirk pair and for the background particles. In \secref{Sec:Tracking}, we discuss the {\sc Exa.TrkX}  pipeline and how it can be applied to quirk track reconstruction. In \secref{Sec:Performance} we present our our track reconstruction performance. In \secref{sec:bgd_sens} we discuss the sensitivity of our reconstruction techniques to quirks. In \secref{Sec:Discussion} we conclude and discuss future directions.

\begin{figure}[h!]
    \centering
    \includegraphics[width = 0.99\textwidth]{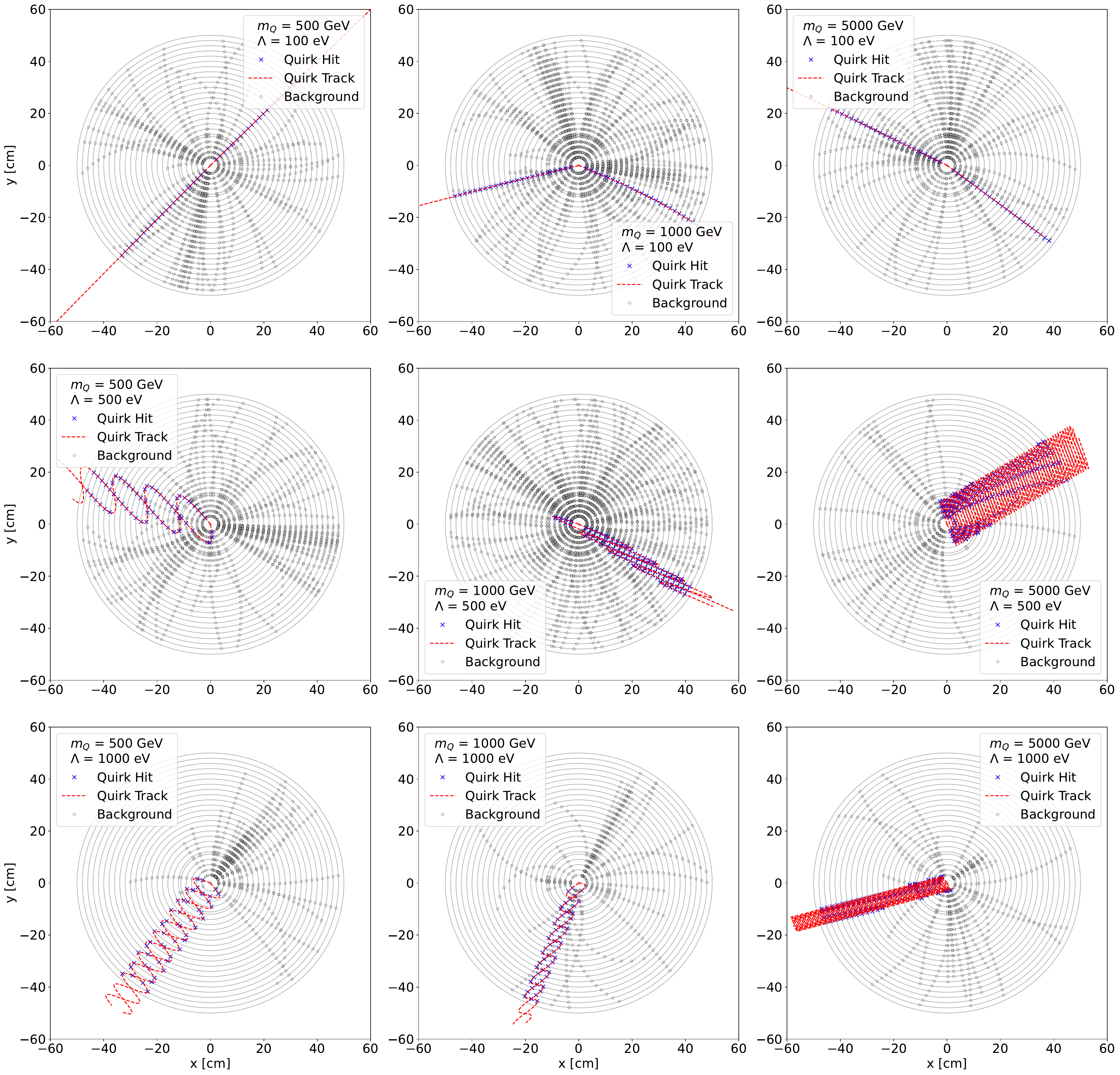}
    \caption{Example trajectories of quirk pairs in simulated events with a simplified detector geometry described in the text. The top row shows example trajectories for $\Lambda$ = 100 eV, and increases in mass from left to right, with $m_Q = 500,1000,5000$ GeV, \rev{all with small oscillation distances}. The center (bottom) row shows examples for the same mass range but for $\Lambda = 500 (1000) $ eV, \rev{both with visible oscillations}. Hits from SM background tracks \rev{(see Sec~\ref{sec:qbg} for details)} are in grey, and from quirks are in blue. The true quirk trajectory is also shown, in red.}
    \label{fig:quirkexample}
\end{figure}

\section{Theoretical Background}
\label{Sec:Background}

In the Standard Model, quarks are charged under the unbroken ${\rm SU}(3)_{\rm C}$ which confines below the QCD scale, $\Lambda_{\rm QCD}$.  Notably, there exists light quarks with a mass $m_q$ such that $m_q \ll \Lambda_{\rm QCD}$. This leads to color confinement, as quarks separated by a sufficient distance convert the energy of their interaction into quark pairs.

In contrast, if the masses were above the confinement scale, macroscopic distances could be maintained. This is the case for {\it quirks}~\cite{Kang:2008ea}, charged under a distinct non-abelian, confining gauge theory ${\rm SU}(N)_{\rm IC}$, sometimes called {\it infracolor}, of which all other standard model particles are singlets. This new gauge interaction produces a confinement scale $\Lambda_{\rm IC}$. The key difference between quirks and quarks is that for a quirk, $Q$, the mass $m_Q \gg \Lambda_{\rm IC}$, where for convenience, we write $\Lambda \equiv \Lambda_{\rm IC}$ for the remainder of this paper. Quirks are also typically charged under the Standard Model, such that they can be pair produced at colliders. 

The unbroken ${\rm SU}(N)_{\rm IC}$ leads to an infracolor gauge string between the pair-produced quirks, as with quarks. Unlike quarks however, as the quirks separate the breaking of the string is exponentially suppressed due to $m_Q\gg \Lambda$. The absence of  string breaking results in oscillation of the pair with a characteristic length~\cite{Knapen:2017kly}
\begin{equation}
    d_{cm} \approx 2 ~{\rm cm} (\gamma-1) \left( \frac{m_Q}{100 ~{\rm GeV}} \right) \left( \frac{{\rm keV}}{\Lambda} \right)^2
    \label{eq:dcm}
\end{equation}
in the center-of-mass frame, where $\gamma$ is the Lorentz boost of the quirks at production. When reporting the quirk track reconstruction performance we will use this formula for qualitative understanding. It is important to note that since we study pair produced quirks in association with a jet, actual oscillation lengths in the lab frame can be very different from Eq.~\ref{eq:dcm} due to smaller-than-typical opening angles or different energies of each quirk. 

In a collider setting, the oscillation length of produced quirks determines the potential reconstruction scenario. For $d_{cm}\lesssim 1$ mm,  oscillation cannot be resolved by the detector, and the quirk pair is reconstructed as a single track. For oscillation scales much larger than the detector, $d_{cm} \gtrsim
 1$ m, the quirk pair does not complete an oscillation within the detector, and so the quirk pair appears as two standard tracks. In both cases, the tracks can be described using helical parameterization,  traditional  tracking algorithms are effective and existing searches are relevant\cite{Knapen:2017kly,Evans:2018jmd, D0:2010kkd, Farina:2017cts}.

However, when the oscillation length is on the order of the size of the detector, the quirk pair  produces two tracks with a unique, non-helical motion. See Fig.~\ref{fig:quirkexample} for visualizations of example trajectories, \rev{whose features are explained in greater detail in the following section}. The stark deviations from a helix make traditional tracking algorithms ineffective for finding such complex tracks.

\section{Modeling}
\label{Sec:Modeling}

Samples of simulated events are prepared and used for training of the tracking algorithm as well as evaluation of the performance. We study quirks  charged under the Standard Model ${\rm U}(1)_{\rm EM}$ \rev{(with charge $\pm1$)} such that they can be pair produced through, e.g., an $s$-channel Drell-Yan process. \rev{Samples of simulated quirk signal and SM background are generated in $pp$ collisions at $\sqrt{s}=13$ TeV with MadGraph5\_aMC~\cite{madgraph}}. Non-quirk particles are showered and hadronized with Pythia8~\cite{Bierlich:2022pfr} and a selection  of $p_{\textrm{T}}>100~{\rm MeV}$ \rev{and} $|\eta|<2.5$ is applied on all charged particles, including quirks.  The trajectories of particles and their interaction with the detector are then modeled with a custom package described below.

\begin{figure}
    \centering
   \includegraphics[width = 0.99\textwidth]{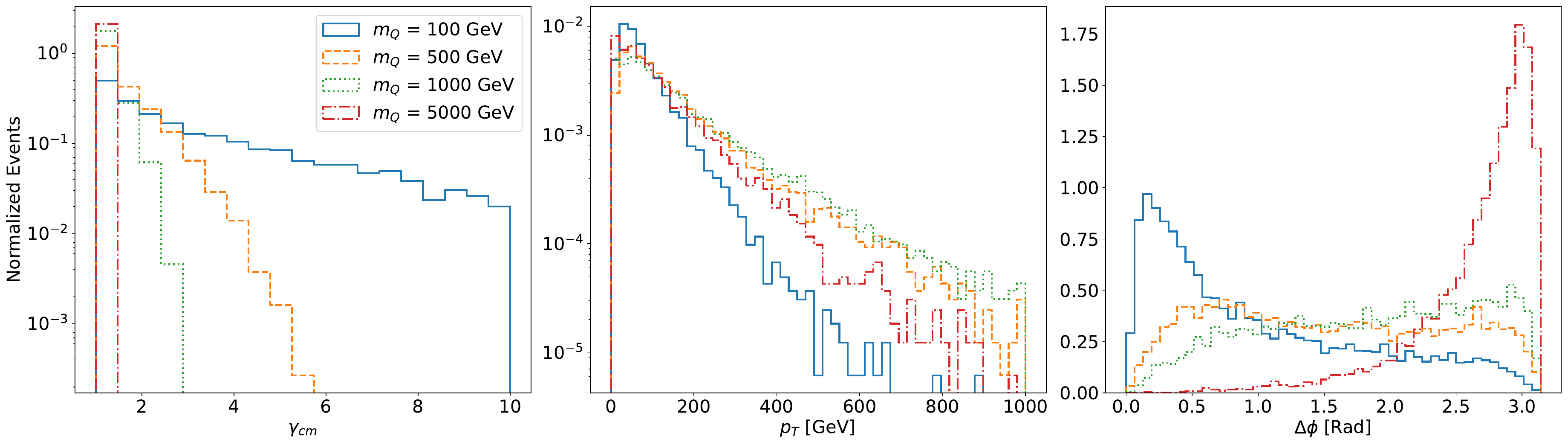}
    \caption{ Distributions of kinematic quantities for simulated quirk pairs, for several mass values, independent of $\Lambda$. Shown are the $\gamma$ factor, transverse momentum $p_\textrm{T}$ and opening angle $\Delta \phi$.}
    \label{fig:prod}
\end{figure}

\rev{The leading order Drell-Yan production mechanism results in a quirk pair with zero transverse momentum, but each quirk can have nonzero transverse momentum.}
 For certain regions of the $(m_Q,\Lambda)$ parameter space with oscillation lengths $\sim 1$ cm, the resultant motion is that the quirks travel in the \rev{beam} direction~\cite{Li:2021tsy} and only leave hits in the innermost layers of the detector. To generate visible tracks, we produce the quirk pair in association with a jet, giving the quirk pair a net transverse momentum. We generate samples of simulated quirks in the mass range $m_Q\in$[100,5000] GeV. Fig.~\ref{fig:prod} shows the distributions of the simulated $\gamma$ factor of the quirk pair, $\gamma_{cm}$, quirk pair transverse momentum and opening angles in this range\rev{; here, the opening angle refers to the initial angle between the quirks within the plane of their production.} These quantities broadly describe the oscillation calculated for quirk trajectories.  In general, as the opening angle, $\Delta \phi$, approaches $\pi$, and for quirks produced with large momenta, the resultant quirk motion will have the largest separation. An additional feature of the initial quirk production which has important consequences for the quirk trajectory is the case where $p_{Q1} \gg p_{Q2}$ (or vice versa). In this case $Q_1$ will drag $Q_2$ in the direction of $\vec{p_1}$ which will result in a complicated motion and exchange of momenta between each quirk.

In $pp$ collisions at $\sqrt{s}=13$ TeV, the production cross section for two quirks and a jet  $\sigma_{QQj}\approx 1 $ fb for $m_Q=500~{\rm GeV}$, which is sufficient to expect a significant number of quirks in the current LHC dataset. For the purposes of this study we will explore quirk masses up to 5 TeV; for the larger masses in this range, the cross section is too small for appreciable production at the LHC with current planned luminosities. Nevertheless, we explore these heavier cases to assess the strength of our reconstruction technique which may be useful at future colliders or for models with higher production cross sections. Larger cross sections can be achieved for quirks with lower $p_\textrm{T}$, with associated reconstruction challenges, or quirks charged under ${\rm SU}(3)_{\rm C}$.

Instead of following a helical path, quirk pairs follow the motion of two objects subjected to a central force. Given that their angular momentum is approximately conserved, their trajectories are largely confined to a single plane\cite{Knapen:2017kly}. Solutions to the equations of motion assume an extra central force between the two quirks in addition to the Lorentz force that arises from the external electric and magnetic fields. Following Ref.~\cite{Knapen:2017kly},  the system is modeled using a Nambu-Goto action with massive endpoints~\cite{Luscher:2002qv}, leading to the equations of motion for a quirk given by the force:

\begin{equation}
\frac{d(m\gamma\vec{v})}{dt} =  -\Lambda^2 \left( \frac{\sqrt{1 - v_{\perp}^2}}{v_{\parallel}} \, \vec{v}_{\parallel} + \frac{v_{\parallel}  }{\sqrt{1 - v_{\perp}^2}}\vec{v}_{\perp} \right) + \vec{F}_{\rm ext},
\end{equation}



\noindent
with a similar equation for the second quirk. Here, $m$ is the mass of the quirk, $\vec{v}_{\perp}$ and $\vec{v}_{\parallel}$ are the perpendicular and parallel velocities of the quirk with respect to the infracolor string, and $F_{\rm ext}$ is the external force acting on the quirks, which for us will only include the affect of the external magnetic field on the (anti-)quirk with charge $q=\pm1$.



To assess the track reconstruction performance, we assume \rev{two} simplified detector configurations for an LHC-style detector. The simplest is an 8-layer~\cite{Knapen:2017kly} configuration with 4 tracking layers in the inner detector between $r=3.1$ cm and $r=12.25$ cm covering $|z|<40~{\rm cm}$  and 4 layers in the outer detector, between $r=29.9$ cm and $r=100$ cm, covering $|z|<75~{\rm cm}$. In a second configuration, we  consider a configuration where there are 25 tracking layers between $r=3.1$ cm and $r=53$ cm \rev{with the same $z$-range as the 8 -layer detector}.  
We calculate particle trajectories and their intersection with a detector layer is recorded as a hit. No noise hits are included in these first studies, though hit-location uncertainties are assessed in \secref{Sec:Uncertainty}.

Figure~\ref{fig:quirkexample}  shows  selected examples of quirk trajectories for $\Lambda = 100, 500, 1000$ eV (top row) and  $m_Q$ = 500 GeV, 1 TeV, 5 TeV. The oscillation length depends strongly on the mass and $\Lambda$, as expected from Eq~\ref{eq:dcm}. As quirks become heavier, they naturally receive less kinetic energy which results in smaller oscillation lengths. For heavy quirks with large $\Lambda$, the oscillations can be quite rapid.  Both very long and very short oscillation lengths are similar enough to traditional helical tracks that standard algorithms can have reasonable efficiency. The focus of this work is on the intermediate regime, where the quirk separation is approximately a few centimeters, making strikingly non-helical tracks.

For the production of background tracks arising from the associated jet production, we collect all stable charged particles on the detector length scale ($e^\pm,\mu^\pm,\pi^\pm,K^\pm$) with $p_\textrm{T}>$100 MeV and $|\eta|<2.5$ from  Pythia. We model their propagation through the  detector as charged particles in a magnetic field, neglecting matter interactions. We find that background tracks contribute many more hits as compared to quirks. For $m_Q\lesssim{\rm TeV}$ we find that background hits outnumber quirk hits by a factor of ${\cal O}(100)$, with lighter quirks being more outnumbered by background than heavier quirks with the same $\Lambda$. For the heaviest quirks we consider, $m_Q~=~5~{\rm TeV}$ which is nearly at the kinematic threshold, there is significantly less energy available for background particles resulting in comparatively fewer background tracks, although background still dominates by a factor of ${\approx}20$.  

\section{Tracking}
\label{Sec:Tracking}

Track reconstruction broadly consists of track {\it finding}, which proposes a set of track candidates given hits in a detector, and track {\it fitting}, which aims to predict the particle momentum and production vector of the matching track candidate. Reconstructed tracks are used for many downstream reconstruction and analysis tasks, including trigger decisions, jet tagging, and signal-background discrimination.

Most traditional algorithms make assumptions about the characteristics of the particles being reconstructed. This particle hypothesis typically includes the assumption that the particle track is approximately helical and that the particle has ionization and interaction properties typical of known particles. In this way, algorithms used in collider tracking such as the kalman filter (KF) propagate a helical (or otherwise modeled) state vector and update it hit-by-hit as part of track finding. Machine learning approaches, such as graph neural networks, need not explicitly model this state and therefore need no such assumption. The characteristics of the particle tracks are learned by providing examples of target tracks, which may therefore be of any shape.

There are a variety of ways ML can be used to solve the track finding problem (see~\cite{duarte2022graph,dezoort2023graph,thais2022graph} for thorough reviews). It can be treated as an object detection problem and one can perform instance segmentation with object condensation~\cite{kieseler2020object,lieret2024object,murnane2024influencer}. One can frame reconstruction as a generative process, where track candidates are recurrently produced by some autoregressive process~\cite{huang2024language}. The combinatorial kalman filter (CKF) technique can also be combined with techniques for extending track seeds using ML~\cite{heinrich2024combined}, or using reinforcement learning to choose a next hit~\cite{vaage2022reinforcement,kortus2023towards}.

A successful approach has been to take advantage of a natural description of tracks as a set of paths in a directed graph, and apply a GNN. The {\sc Exa.TrkX} pipeline is built on this GNN paradigm, and uses the physics intuition of a track being a chain of sequentially connected hits~\cite{farrell2018novel,ju2020graph,ExaTrkX:2021abe}. This and similar approaches have been shown to give high track-finding efficiency and stable training on a diverse set of HEP experiments~\cite{caillou2022atlas,verma2020particle,hewes2021graph,drielsma2021scalable,akram2022track,jia2024besiii,biscarat2021towards,zdybal2024machine,andrews2021accelerating,dezoort2021charged,Correia_2024,Elabd_2022}. This pipeline contains three stages: a) Graph construction, where 3D spacepoints are connected into a graph of spacepoint-to-spacepoint edge connections; b) Edge classification, where edges are scored with a confidence of being either ``true" (a connection between two sequential hits from a signal particle) or ``fake" (any other kind of edge); and c) Graph segmentation, where low-scoring edges are removed, and collections of connected hits are used to produce track candidates.

The {\sc Exa.TrkX} pipeline can be applied nearly out-of-the-box to non-helical tracking, as in the case of quirk tracks. Some alterations to the definition of ``true edges" generalize the pipeline to train correctly for this broader class of tracks. The most significant alteration is in the assumptions about the ordering of the hits. As many tracking detectors (as in ATLAS and the open detector used in the TrackML challenge \cite{Kiehn:2019tbl,Amrouche_2019,
amrouche2021trackingmachinelearningchallenge}) do not include reliable timing information, the ordering of hits in tracks within {\sc Exa.TrkX} is defined heuristically using increasing distance from the production vertex of the track. This is reliable for high-momentum helical tracks that are unlikely to scatter back towards the interaction point. However, quirk tracks with a non-negligible opening angle are able to oscillate away from and towards the detector origin. Relying on the assumption that hits are in order according to their distance from the production vertex can lead to misinterpretation of the true track shape in training. In this work, we access the underlying simulation data to define the time-ordering of hits in order to correctly construct true edges for training, though such information is not required for testing.

To construct graphs, we train a \textit{metric learning} model, which attempts to embed sequential hits in a particle track into nearby positions in a latent space, while hits from other particles should be distant in the latent space. As this is purely data driven, tracks need not be helical, and any smooth particle trajectory may be learned by such a non-linear embedding. All that is required is a definition of pairwise similarity, as defined earlier. 
\rev{In our training analyses, we split our dataset into training, validation, and testing sets with 2k, 400, and 400 samples respectively.}
We use 150 training epochs and a learning rate of 0.001 in the metric learning for constructing the tracking graph. We then train a GNN to classify the edges as either ``true" or ``false". In this step, we modify the epochs and learning rate for better efficiency and to avoid overfitting \rev{by checking the loss and efficiency plots}, using early stopping and decreasing the learning rate. \rev{Also, we define the ``score cut" to select our target candidates. This value is set to 0.9 in most analyses and is optimized in some cases.}

\section{Track Reconstruction Performance}
\label{Sec:Performance}

Performance of the tracking algorithm is measured in simulated samples described above.  Performance is measured by tracking efficiency, the fraction of {\it reconstructable} particles that are {\it reconstructed}. Reconstructable particles are those with at least 7 (22) true hits in the 8-layer (25-layer) geometry. A particle is {\it reconstructed} if it is double-majority matched\footnote{Double-majority matching requires that at least 50\% of the reconstructed track hits belong to a matching truth track, and that at least 50\% of the hits in that truth track are found in the reconstructed track.} to a \textit{matchable} track candidate: one with least 5 (15) hits in the 8-layer (25-layer) geometry.

Below, we demonstrate the challenges of finding quirks using helical-track finders, study the power of ML tracking to find quirk tracks and explore the ability of the pipeline to generalize from one set of quirk parameters to another. Rather than provide an exhaustive set of results for every configuration discussed below, we begin with a single  characteristic choice of quirk model parameters, our {\it benchmark sample}, with $m_Q=500$ GeV and $\Lambda=500$ eV, before presenting a broader study across all parameters. We choose this benchmark as it displays an oscillation length in the region of interest ${\cal O}({\rm cm})$.

\subsection{Reconstructing quirk tracks using SM tracks}

We wish to compare the performance of a traditional helix-finding algorithm to a dedicated ML quirk-finding approach. However, while powerful, the combinatorial Kalman filter is costly to tune on a new geometry. As such, for this study we take a GNN trained on helical SM tracks as the baseline, as it achieves similar physics performance as non-ML algorithms \cite{Torres:2876457}. 

The  SM-trained GNN-based tracker excels at finding SM tracks. For our benchmark sample, in the 8-layer geometry, SM tracks are found with 97.9\% efficiency, while quirks are mostly missed, with 10.2\% efficiency, using the definition of track efficiency above. For the 25-layer geometry, we find similar performance.
 
These results are typical for other $m_Q,\Lambda$ in the regime of interest, where the oscillation length is centimeters. As expected, if the oscillation length is much longer, these trackers succeed to reconstruct the tracks due to the quirk track in the longer oscillation is similar as the SM tracks; See Table~\ref{tab:eff_lambda_Mq_forSM} in Appendix~\ref{ref:app_smeff} for details. 

\subsection{Well-behaved quirks}

Tracks of quirks are substantially more complex than helical tracks, and over the range of parameters can display a broad range of behavior that can be challenging for reconstruction algorithms. As seen in Fig.~\ref{fig:quirkexample}, tracks can cross individual layers multiple times due to oscillations. Before tackling these more difficult scenarios, we focus on the subset of quirk tracks  identified as {\it well-behaved}, those whose number of hits do not exceed the number of layers by a factor of three. This eliminates, for the moment, the most challenging quirk tracks, those which leave a very large number of hits as they travel in and out of detector layers, but retains tracks with clearly non-helical trajectories.

\begin{figure}
    \centering
    \includegraphics[width = 0.45\textwidth]{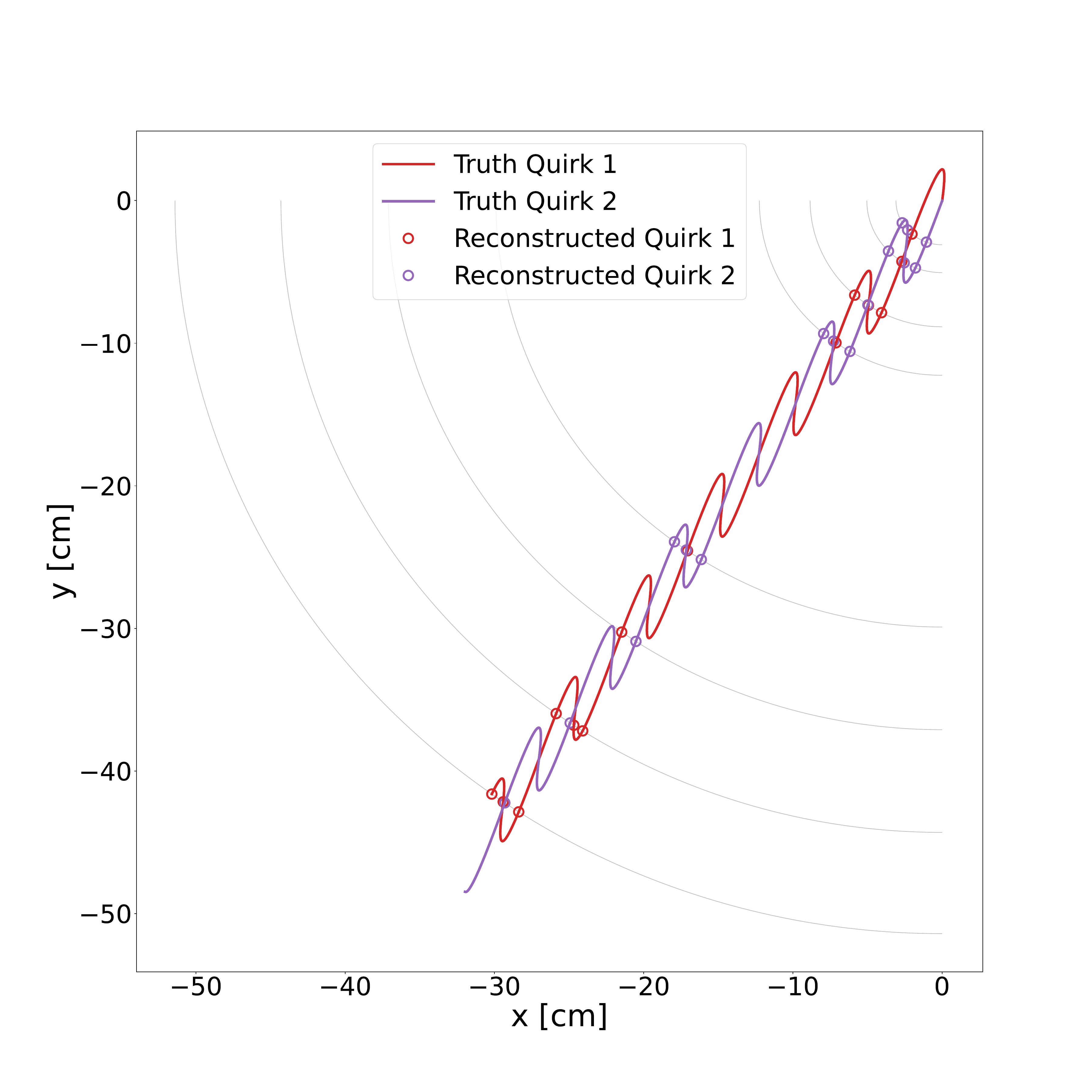}
     \includegraphics[width = 0.45\textwidth]{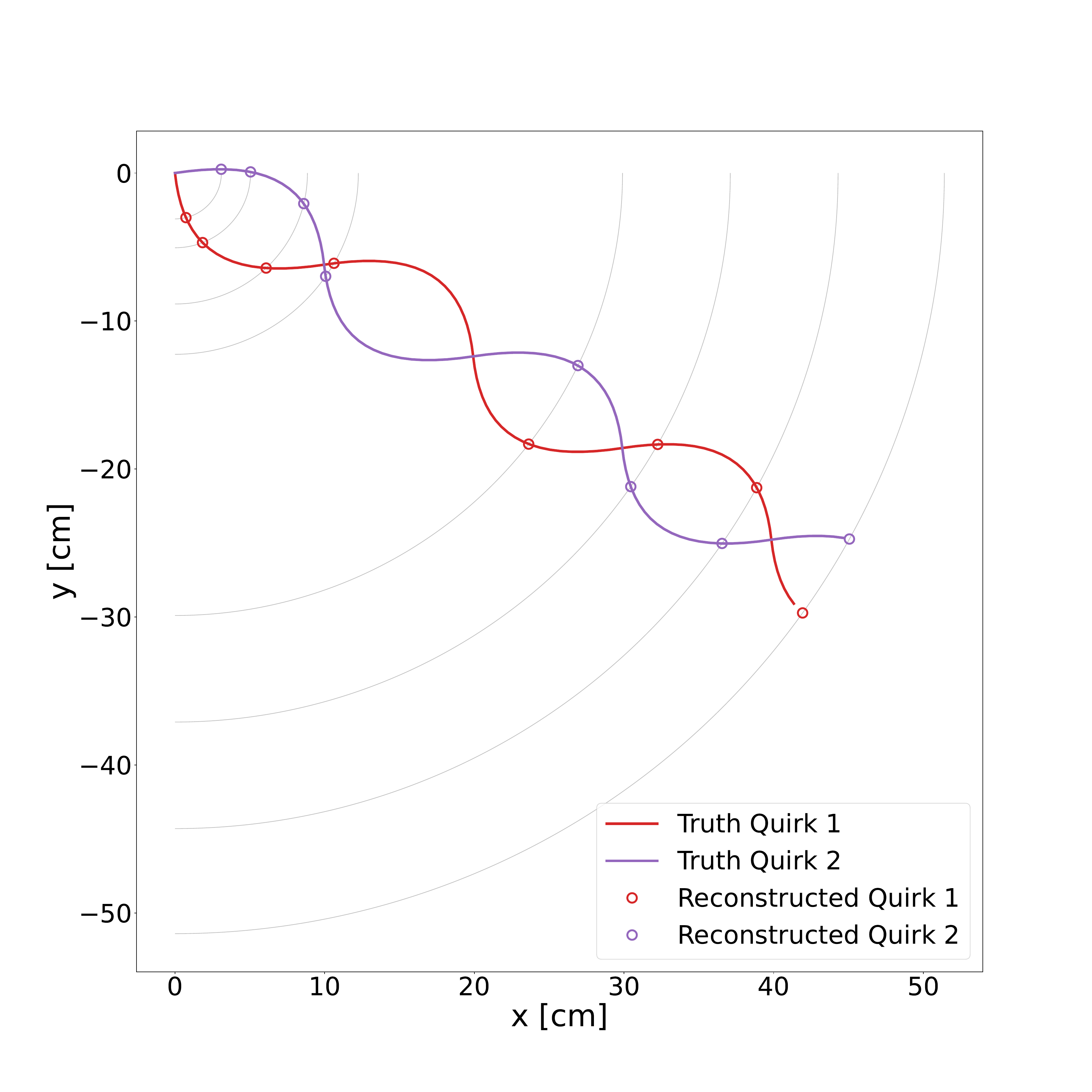}
    \caption{ Example truth and reconstructed well-behaved quirk tracks with $(m_Q=500$ GeV, $\Lambda=500$ eV).}
    \label{fig:well_behaved_quirk_performance}
\end{figure}

In the benchmark sample with 8 layers,  training on a pure sample of well-behaved quirks produces a track finder with efficiency of 91.5\% in events with only well-behaved quirks and no background tracks. This is a highly idealized scenario, but gives a rough upper bound on the performance of the pipeline and demonstrates its capacity to learn to find non-helical tracks when they comprise its training set, without background. Figure~\ref{fig:well_behaved_quirk_performance} shows an example of reconstructed and truth tracks in this scenario.

\subsection{All quirks}

\begin{figure}
    \centering
    \includegraphics[width = 0.45\textwidth]{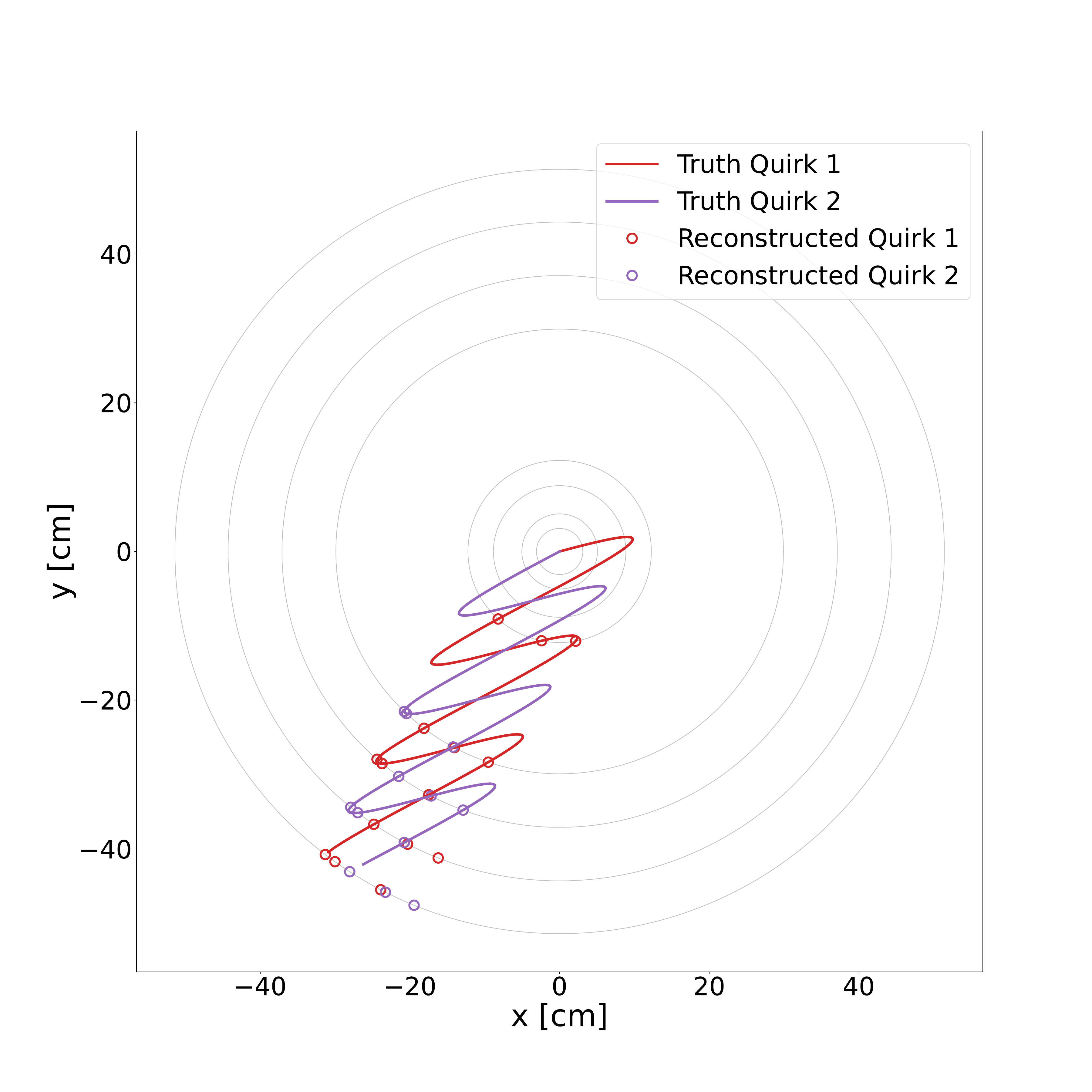}
     \includegraphics[width = 0.45\textwidth]{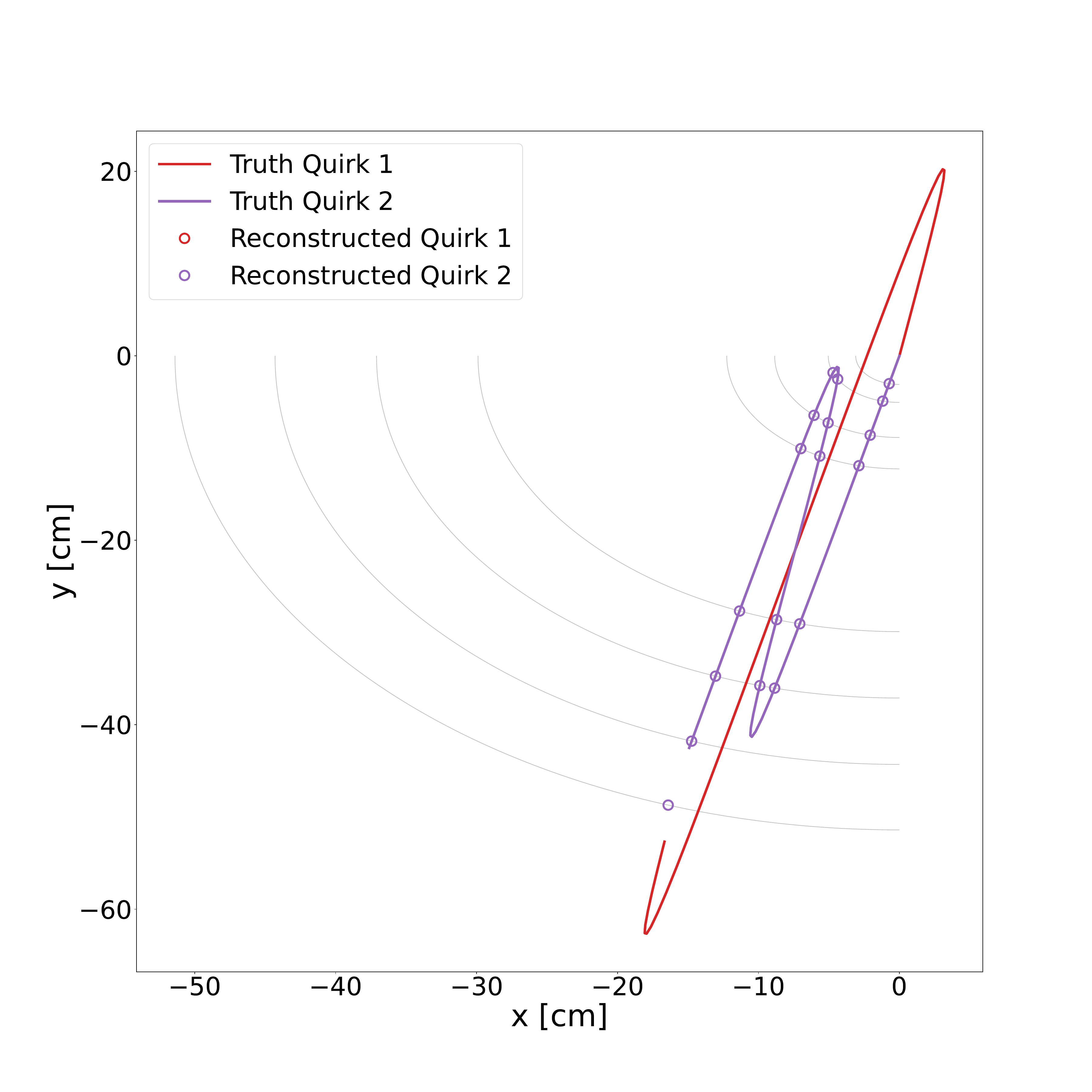}
    \caption{Examples of truth and reconstructed tracks for not well-behaved quirks with $m_Q=500$ GeV, $\Lambda=500$ eV.
    }
    \label{fig:all_quirk_performance}
\end{figure}

Quirks are an interesting and challenging problem precisely because they are not always well-behaved, and performance on these not-well-behaved tracks is also important for broad sensitivity. When the training sample is expanded to include all reconstructable tracks, the efficiency in our benchmark scenario with 8 detector layers is  $56.3\%$ as compared to 10.2\% for the tracker trained on SM tracks. Figure~\ref{fig:all_quirk_performance} presents an example of reconstructed and truth tracks. 

\begin{figure}
    \centering
    \includegraphics[width = 0.32\textwidth]{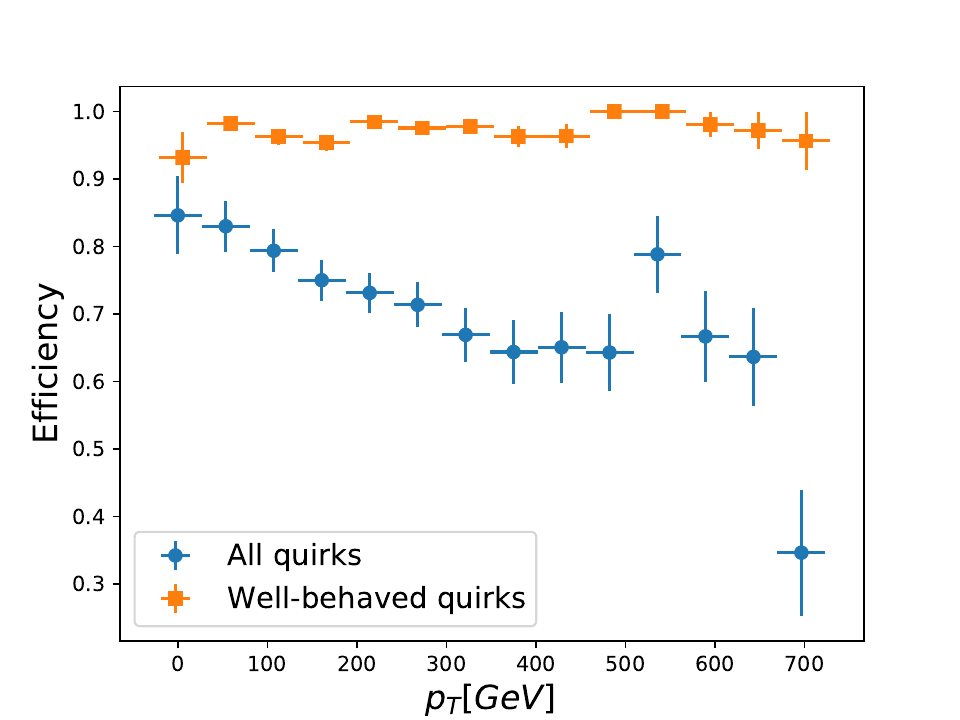}
    \includegraphics[width = 0.32\textwidth]{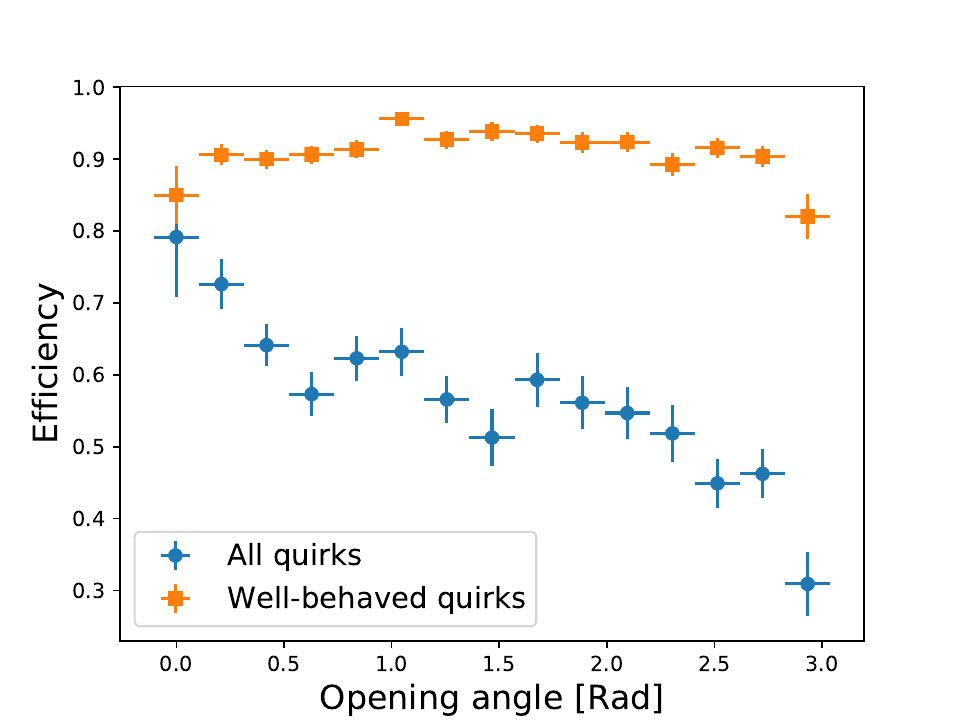}
    \includegraphics[width = 0.32\textwidth]{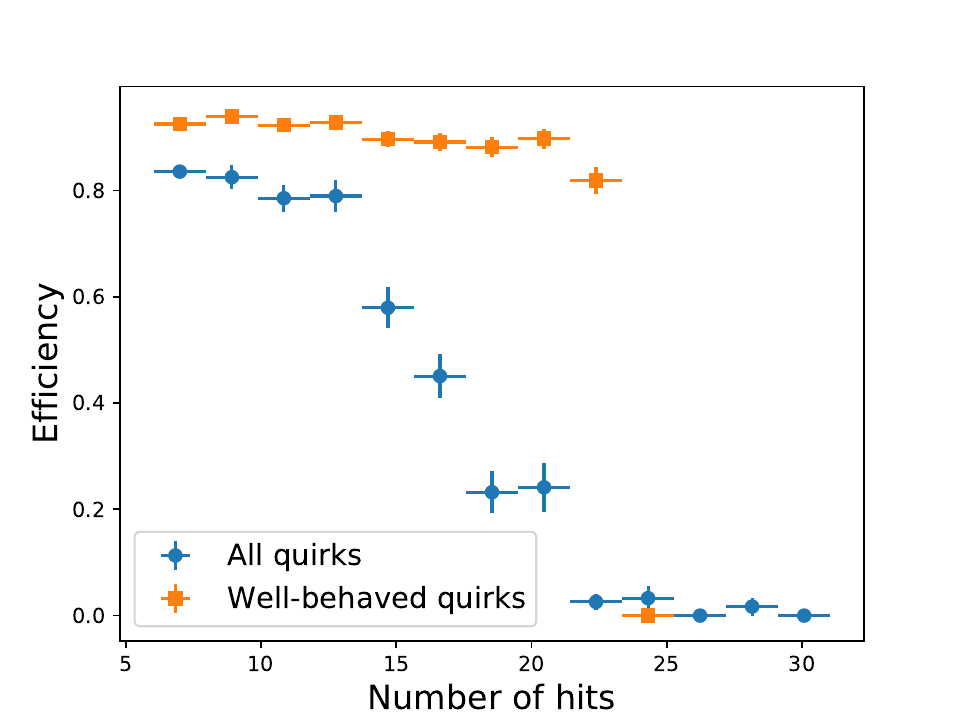}
    \caption{Tracking performance for quirks without background for $m_Q=500$ GeV, $\Lambda=500$ eV in the eight-layer geometry.  Shown is efficiency versus track $p_\textrm{T}$ (left),  quirk opening angle (center), and number of true hits (right) for well-behaved quirks or all quirks. The error bars indicate statistical errors. }
    \label{fig:quirk_perf}
\end{figure}

The efficiency with respect to $p_\textrm{T}$, opening angle and the number of true hits are shown in Fig.~\ref{fig:quirk_perf}, for all quirks as well as the well-behaved subset.  The efficiency falls as the number of hits increases, reflecting the challenge of reconstructing tracks with complex in-and-out trajectories.
There is also a soft dependence on opening angle and quirk track $p_\textrm{T}$, due to tracks with wider opening angles or higher momentum being more likely to exhibit complex oscillations with in-and-out behavior.  

The performance on the full set of quirk tracks is a significant improvement relative to the efficiency of the SM-trained algorithm. However, due to the complexity of reconstruction of the all-quirks dataset,  we leave the broader challenge of optimizing the all-quirks reconstruction, and indeed any parameterizable path, to future work, and base our analysis on  well-behaved quirks in the following sections. The well-behaved quirks contain sufficient complexity to represent a novel tracking challenge and a large enough fraction of the all-tracks dataset to provide sensitivity to quirks at the LHC. 

\subsection{Finding Quirks among the Background}
\label{sec:qbg}

Realistic applications of non-helical tracking  require identifying such tracks among many background helical tracks from SM processes. To assess the performance in this environment, we use a training sample which includes tracker hits from both quirk and SM particles in each event. In graph construction and edge classification, only quirk hit pairs are treated as true. All other pairs, including pairs of hits from the same SM particle, are treated as fake in training, and models are encouraged to \textit{not} reconstruct them. Previous work has explored other strategies with such background tracks, such as treating them as true or masking them from the loss function \cite{caillou2022atlas}. In this work, our goal is quirk finding at the highest possible background rejection rate, rather than the reconstruction of all true tracks, and this motivates the strategy as described. The efficiency in this scenario is $71.9$\% for all quirk tracks in our benchmark sample with eight detector layers, somewhat less efficient than the much simpler task of background-free tracking, which scores 91.5\%.  \rev{Background tracks are also included in training discussed in subsequent studies in the following sections, as well as the results in Tab.~\ref{tab:eff_lambda_Mq}.}

\subsection{Dependence on Geometry}

\begin{figure}
    \centering
    \includegraphics[width = 0.45\textwidth]{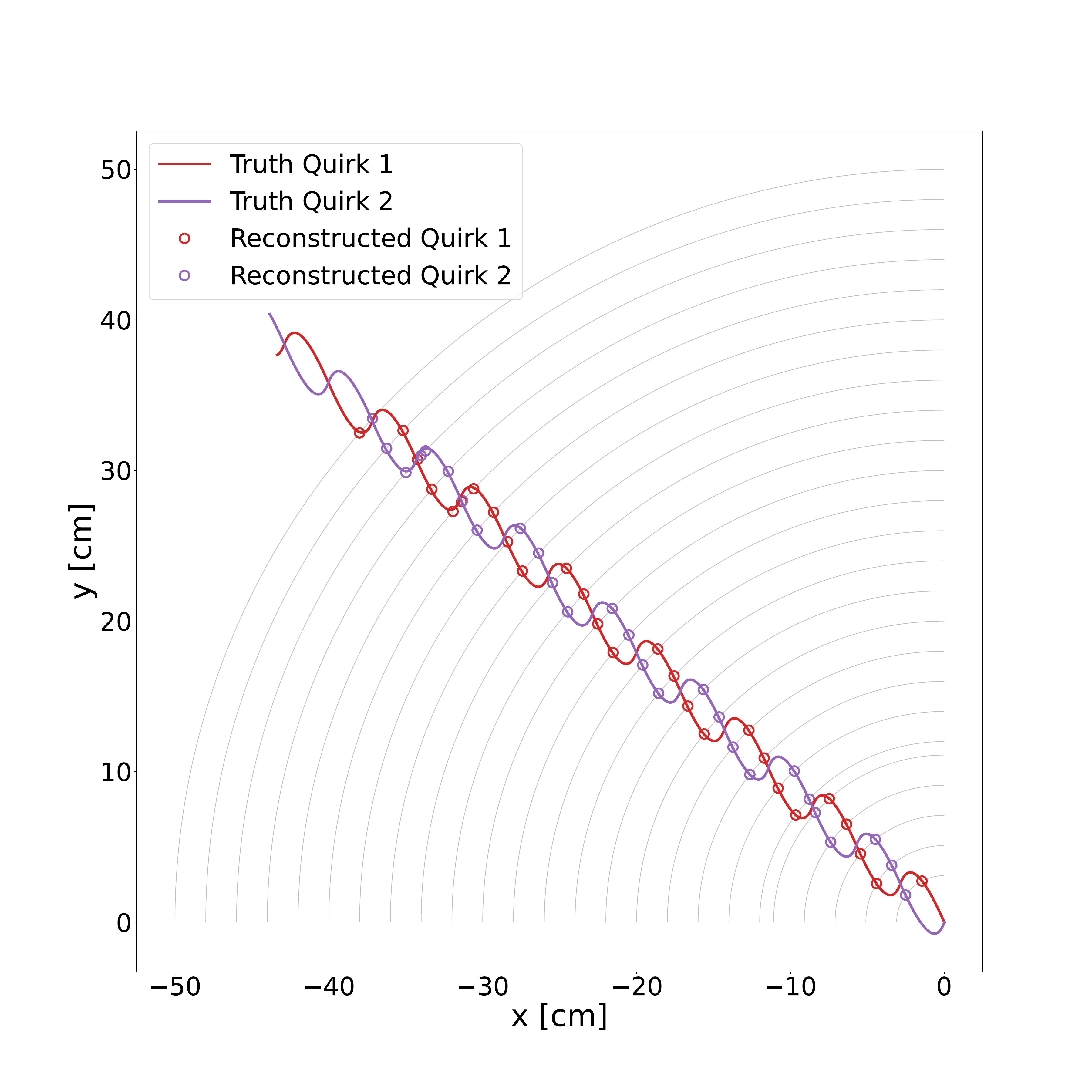}
     \includegraphics[width = 0.45\textwidth]{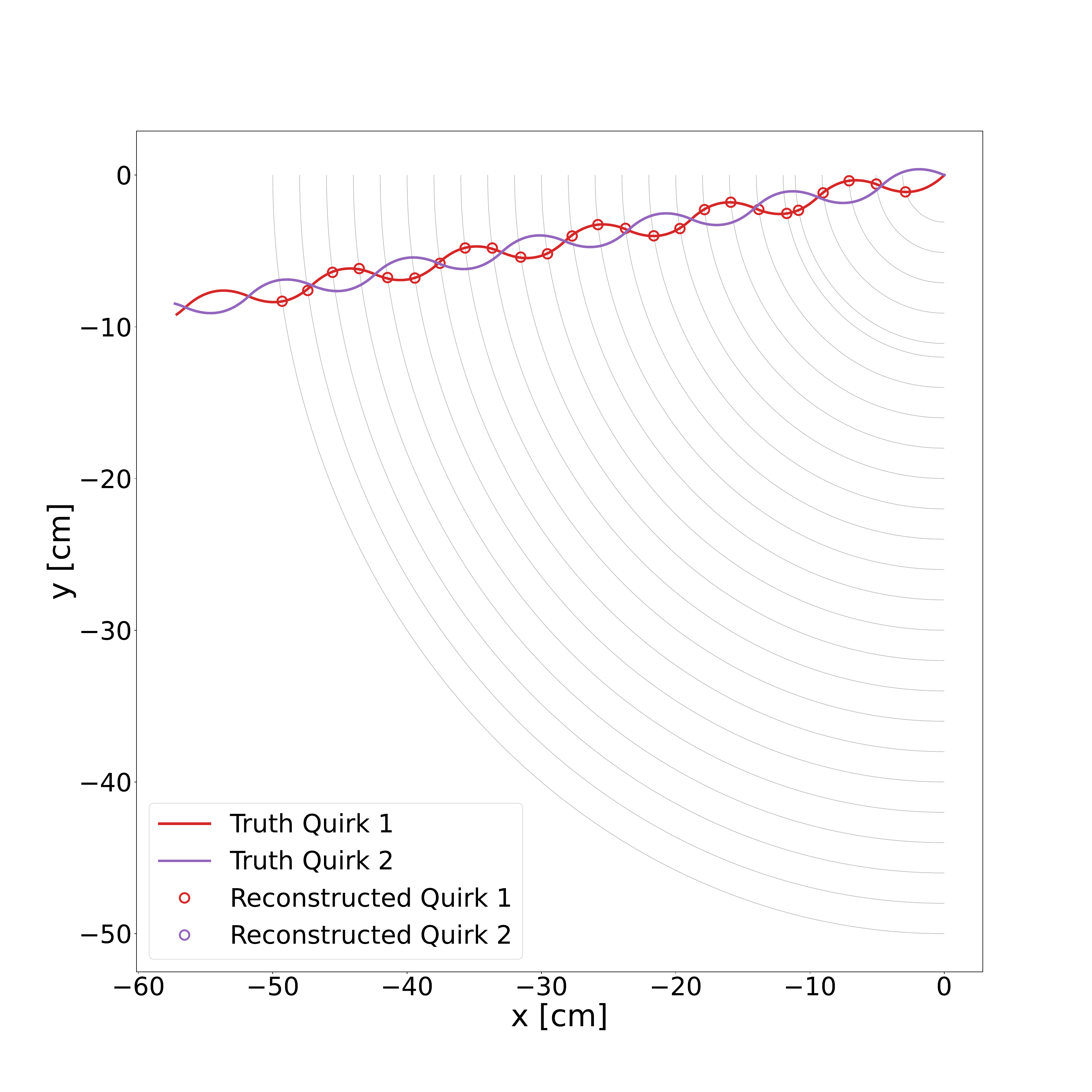}
    \caption{Example truth and reconstructed quirk tracks for $m_Q=500$ GeV, $\Lambda=500$ eV in the 25 layer geometry. The "X" represents truth hits.}
    \label{fig:mix_performance_m500_L500}
\end{figure}

To explore the dependence of the number of tracking layers, the study was repeated in the 25-layer setting, yielding an efficiency of $79.0\%$, somewhat higher than the efficiency in the 8-layer case. This suggests that the nature of the problem does not sensitively depend on the number of layers. Figure~\ref{fig:mix_performance_m500_L500} presents an example of reconstructed and truth tracks with 25 layers for $m_Q=500$ GeV, $\Lambda=500$ eV. \rev{The 25-layer geometry is used in studies in the following sections.}


\subsection{Dependence on Quirk Parameters}

Results above demonstrate effective tracking for a single value of quirk parameters, but the task of quirk track reconstruction is much broader, and the nature of the tracking challenge varies strongly with the parameter values. Table~\ref{tab:eff_lambda_Mq} shows the typical oscillation length and measured reconstruction efficiency for  quirks over a range of quirk parameters $m_Q$, $\Lambda$.

\begin{figure}
    \centering
    \includegraphics[width = 0.45\textwidth]{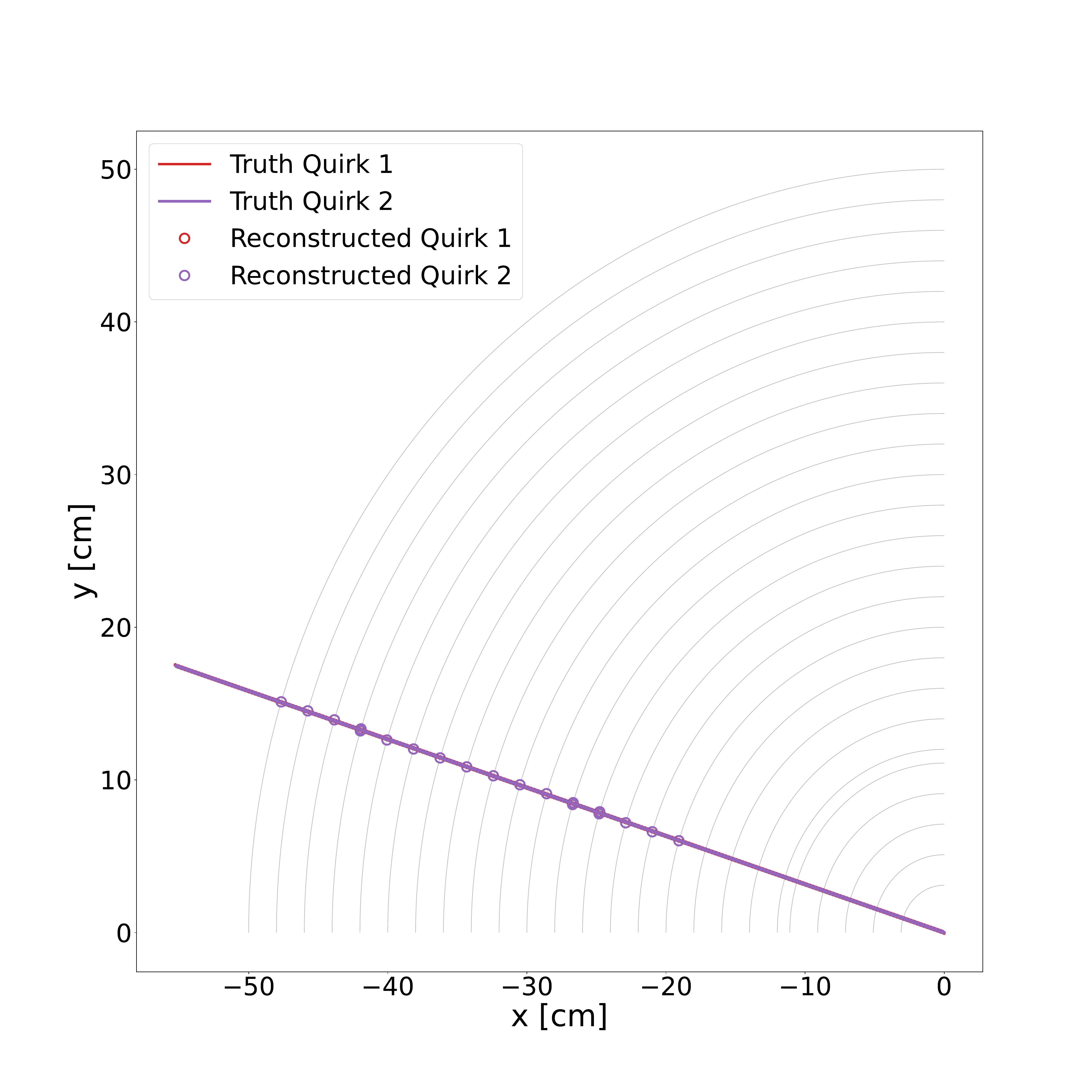}
    \includegraphics[width = 0.45\textwidth]{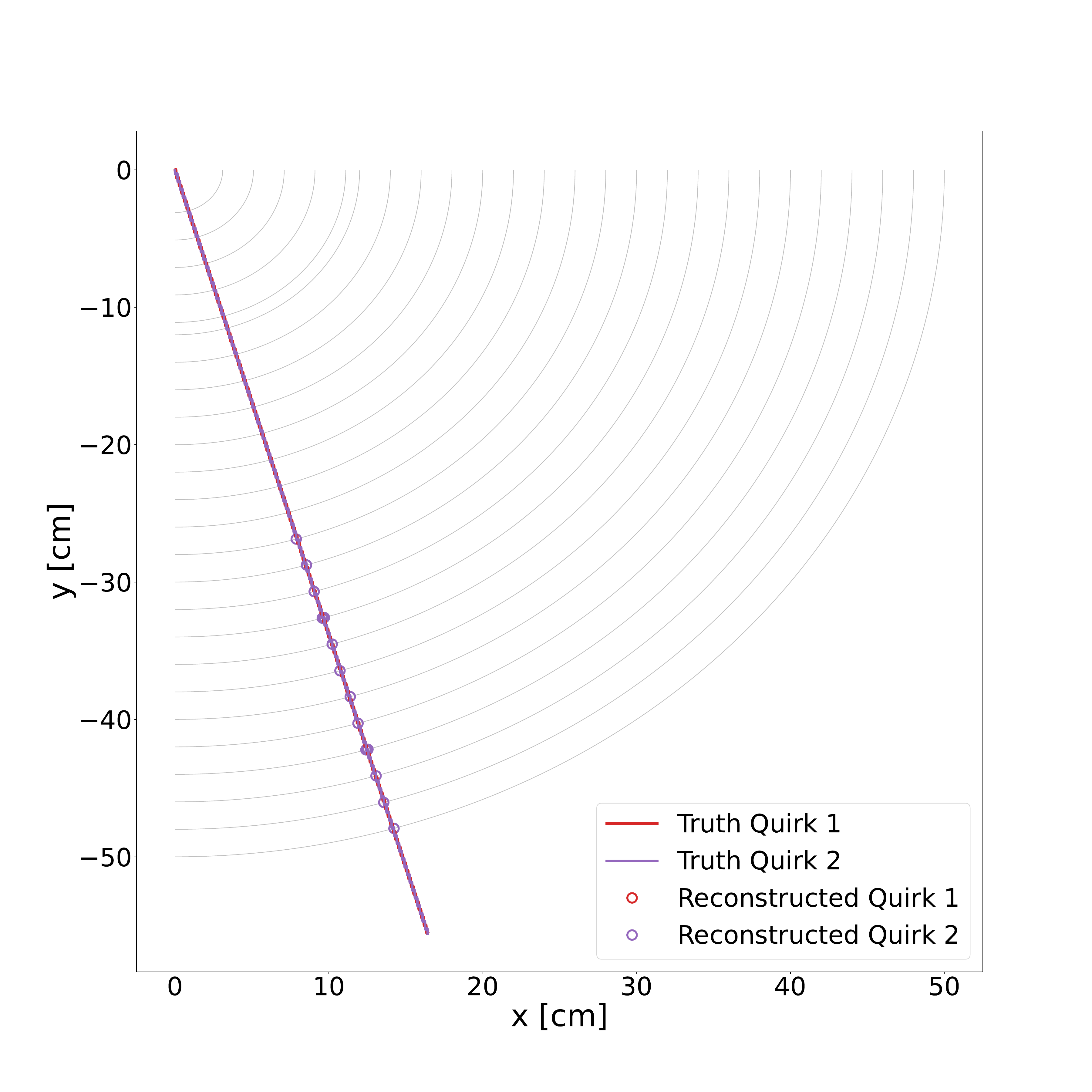}
    \caption{Examples of truth and reconstructed tracks for well-behaved quirks with $m_Q=500$ GeV, $\Lambda=5000$ eV, in which there is no visible oscillation and the two quirks overlap. The "X" represents truth hits.}
    \label{fig:all_quirk_noosc}
\end{figure}

 For very long oscillation lengths, ${\cal O}({\rm m})$, no oscillations occur in the tracking volume and the problem is essentially indistinguishable from that of the SM tracking, and efficiencies are very high, $\approx90$\%.  For the target regime of intermediate-length oscillation lengths, ${\cal O}({\rm cm})$, performance is in the range of $60-80$\%, which is a dramatic improvement over the $\approx 10
 \%$ efficiency for quirk tracks when using a SM-trained tracker (compare Tab.~\ref{tab:eff_lambda_Mq} Tab.~\ref{tab:eff_lambda_Mq_forSM}).  As the oscillation length becomes very small, reported efficiency drops by a factor of $\approx2$; this is due to the two quirk tracks overlapping nearly exactly, often leading to only one reconstructed track, though in principle two exist; see Fig.~\ref{fig:all_quirk_noosc}.

 \subsection{Generalization across Quirk Parameters}

The studies above demonstrate the capacity for the pipeline to reconstruct tracks for particular quirk parameters when those tracks are present in the training sample. It would be much more desirable, however, to develop a {\it single} pipeline capable of reconstructing {\it any} quirk track, or more ambitiously, any smooth, continuous particle path. Achieving such a broad capacity is left to future work, but the rough ability of the current pipeline to generalize can already be assessed by training on a sample with several  quirk parameter sets, rather than a single choice, and testing on a sample of tracks from a parameter set which does not appear in the training sample.   

The efficiency to reconstruct quirks generated with $m_Q=100$ GeV, \rev{$\Lambda=1000$ eV} is 52.6\% when training with a mixture that does not include that set, compared to 77.8\% when training only on tracks from the same parameter set as the testing sample. \rev{The mixed set includes $m_Q~({\rm GeV}),\Lambda~({\rm eV})=(100,100), (500,100), (1000,100) (100,500) (5000,1000)$}. This demonstrates some significant but imperfect ability to generalize. Another working point with $m_Q=1000$ GeV, $\Lambda=1000$ eV achieves 49.3\% efficiency for the generalized training, compared to 62.7\% for the specific training, a similar performance ratio. The nature of the tracking challenge varies with quirk parameter values, but each working point clearly has enough in common for the tracker to learn general features of quirk tracks, not just the features of quirks for one particular quirk parameter choice. Further work is required to understand which ML choices contribute to the highest generalizability.

\subsection{Hit Location Uncertainty}
\label{Sec:Uncertainty}

To explore the sensitivity of the pipeline to imperfect detector resolution, we smear the hit locations \rev{in training and testing} with a Gaussian of width 1 mm, which is larger than typical LHC tracking resolution~\cite{Mindur:2017nqn, CMS:2014pgm}. A spot test for $m_Q=500$ GeV, $\Lambda=500$ eV quirks shows no significant degradation in tracking performance, with performance consistent within statistical uncertainties. 


\begin{table}[]
    \centering
     \caption{ Efficiency to reconstruct quirk tracks for various values of quirk mass ($m_Q$), confinement scale ($\Lambda$) based on 25-layer and {\it well-behaved} setting. The {\it well-behaved} fractions for each point are shown below.  We also calculate the mean Lorentz factor of the quirks ($\bar{\gamma}$) and the standard deviation ($\sigma_\gamma$) in each of the generated datasets to calculate oscillation length $d$ as calculated by Eq.\ref{eq:dcm} which can be useful for understanding reconstruction efficiencies. See text for details.}
    \label{tab:eff_lambda_Mq}
    \begin{tabular}{cccccc|c}
    \hline \hline
         $m_Q$ (GeV)  & $\Lambda$(eV) &$\bar{\gamma}_{\rm }$ & $\sigma_\gamma$ & $d$ [cm] &  Efficiency & {\it Well-behaved} fraction\\
         \hline
         100 & 100 & 4.4 & 3.4& 670 & 91.0\%    & 88.3\%\\
          & 500  & 3.7 & 2.7 & 21.6 & 82.8\%    & 77.0\%\\
          & 1000&  3.1& 2.05 & 4.2  & 77.8\%    & 79.7\%\\
          & 2000& 3.0 & 2.0 & 1.0   & 60.4\%    & 83.4\%\\
          & 3000& 3.0 & 1.9 & 0.44  & 35.6\%    & 83.6\%\\
          & 4000& 2.9 & 1.8 & 0.24  & 34.5\%    & 84.5\%\\
          & 5000& 2.9 & 1.7 & 0.15  & 24.2\%    & 85.0\%\\
         \hline
         500 & 100& 1.9 & 0.7 & 896 & 92.0\%    & 82.3\%\\ 
          & 500&1.8 &0.6 & 31       & 79.0\%    & 51.2\%\\
          & 1000&1.7 &0.6 & 7.3     & 64.3\%    & 53.1\%\\
          & 2000&1.7 & 0.5& 1.7     & 60.9\%    & 59.9\%\\
          & 3000&1.7 &0.5 & 0.8     & 59.6\%    & 62.3\%\\ 
          & 4000&1.6 &0.5 & 0.4     & 42.6\%    & 63.0\%\\
          & 5000&1.6 &0.5 & 0.2     & 39.2\%    & 63.8\%\\
                  \hline
         1000 & 100&1.5 &0.3 & 950  & 94.6\%    & 80.5\%\\
          & 500& 1.4& 0.3& 32       & 63.6\%    & 40.2\%\\    
          & 1000&1.4 &0.3 & 7.6     & 62.7\%    & 42.3\%\\
          & 2000&1.4 &0.3 & 1.8     & 69.2\%    & 48.6\%\\
          & 3000& 1.3&0.2 & 0.8     & 54.1\%    & 51.7\%\\
          & 4000& 1.3&0.2 & 0.4     & 59.7\%    & 52.9\%\\ 
                  \hline
         5000 & 100&1.04 &0.03 & 420    & 84.8\%    & 40.2\%\\
          & 500&1.03 &0.02 & 11         & 69.8\%    & 32.6\%\\     
          & 1000& 1.03& 0.02& 2.7       & 69.5\%    & 35.2\%\\  
          & 2000& 1.03 & 0.02 & 0.7     & 49.2\%    & 39.6\%\\
          & 3000& 1.03& 0.02& 0.3       & 36.4\%    & 40.8\%\\
          & 4000&1.03 & 0.02& 0.2       & 34.6\%    & 41.2\%\\
         \hline
         \hline
    \end{tabular}
\end{table}

\section{Backgrounds and Sensitivity}\label{sec:bgd_sens}

\begin{figure}[h!]
    \centering
    \includegraphics[width=0.4\linewidth]{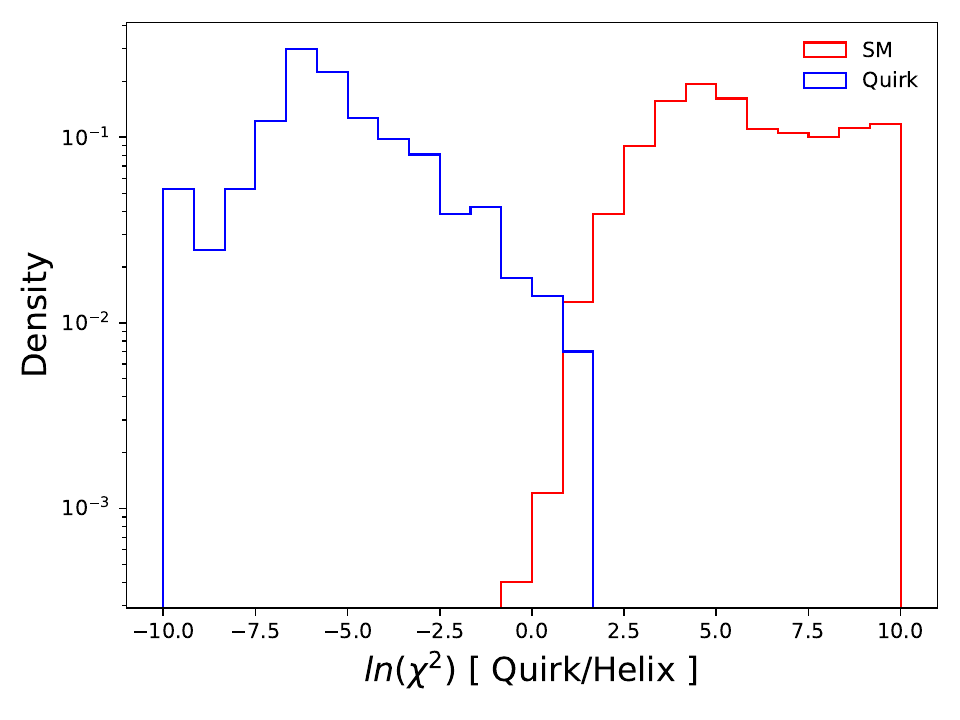}
    \includegraphics[width=0.4\linewidth]{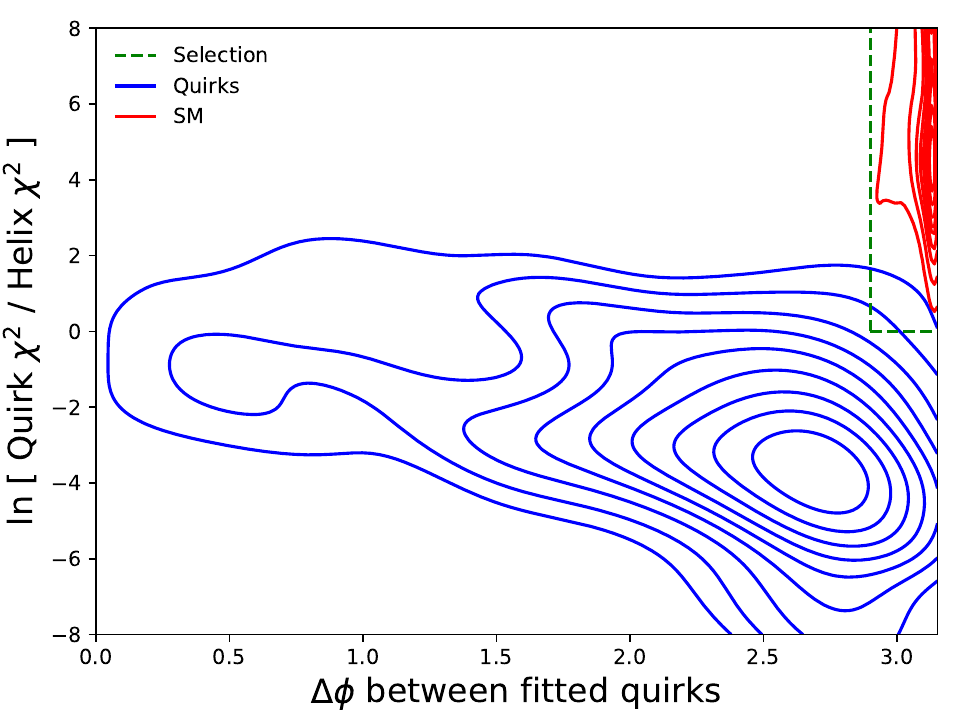}
    \caption{Distributions of ratio $\chi^2$ under the quirk and helix hypothesis, for $m_Q=500~{\rm GeV},\Lambda=500~{\rm eV}$quirk and SM tracks. The lower panel shows the ratio of the two $\chi^2$ metrics versus the fitted quirk opening angle for the same $m_Q,\Lambda$, and the selection used to remove SM tracks is shown in the dashed red lines.}
    \label{fig:chisq}
\end{figure}

\begin{figure}
    \centering
    \includegraphics[width=0.9\linewidth]{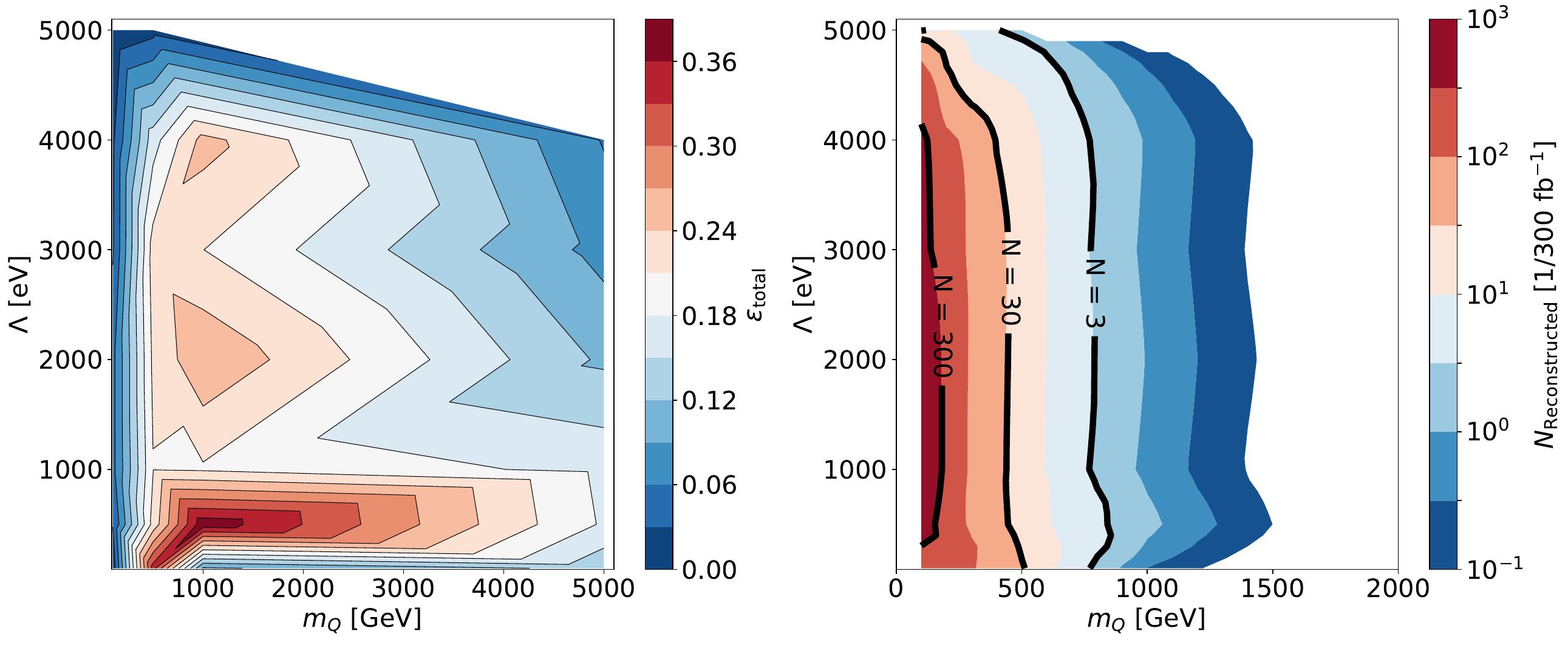}
    \caption{Left: Efficiency of quirk finding versus mass $m_Q$ and string tension $\Lambda$. Right: Expected reconstructed quirk event rate 
 per 300 fb$^{-1}$ in the same parameter plane.}
    \label{fig:quirkeff}
\end{figure}

\rev{
The pipeline has demonstrated the capacity to find quirks with a reasonable efficiency, which is necessary to observe quirks in experiments at the LHC. However, equally important to control is the background (or "fake") rate. Quirks are an exciting target because their signature is striking enough to suggest the possibility of very low background, potentially even background-free, discoveries. Unfortunately, SM tracks have a very high rate, so even a tiny probability to mimic the quirk signature could significantly impact discovery potential.

To examine the rate of background from SM tracks, we collect samples of fake tracks reconstructed by the pipeline which are not matched to quirks.  We perform two fits to each track, one to the quirk hypothesis and a second to the helix hypothesis and use the ratio of $\chi^2$ values to discriminate between quirks and the background.  In both the quirk and helix hypothesis, the $\chi^2$ is the sum of the squared distances between a hit and the distance of closest approach by the track, with an implicit $\sigma^2=1$~cm$^2$ for simplicity.

The helix fitter finds the value of the five helical parameters that minimize the $\chi^2_{\textrm{helix}}$, starting from an initial estimate~\cite{HANSROUL1988498} and refined by the Nelder-Mead optimization algorithm. This approach does not match the sophistication of track fitting in a realistic setting, which benefits from additional heuristics, such as outlier removal and ambiguity resolution. For our purposes, this provides a conservative estimate on our ability to distinguish between true quirks and background. The distribution of $\chi^2_{\textrm{helix}}$ is shown in the upper left panel of Fig.~\ref{fig:chisq} for true quirks \rev{at our benchmark point} $(m_Q=500,\Lambda=500)$ and for background particles.  There is a clear separation between the two distributions, as many quirks cannot be well fit to a helix.  A significant population of background tracks also receive a $\chi^2_{\textrm{helix}}$ larger than what might be expected, due to incorrect hits and local minima in the fitting.

In the quirk hypothesis, we use the propagation algorithm as described in Sec.\ref{Sec:Modeling}, which requires that we fit simultaneously for the four-vectors of the quirk pair given values of $(m_Q,\Lambda)$, though only one track is required and included in the $\chi^2$ evaluation. The distribution of the ratio of $\chi^2_{\textrm{quirk}}$ to $\chi^2_{\textrm{helix}}$ values are shown in the left panel of Fig.~\ref{fig:chisq} for true quirks and for background SM tracks.  Both  background tracks can be accomodated by the quirk hypothesis, due to the freedom provided by the second quirk's unconstrained parameters. To mimic a high-momentum SM track, the fitter finds solutions in which the two quirks are back to back, generating perfectly straight line.  To suppress these, we remove tracks whose fitted quirks are back to back ($\Delta \phi>2.8$) or whose ratio of $\ln({\chi^2_{\textrm{quirk}}/\chi^2_{\textrm{helix}}}) < -1$; see the lower panel of Fig.~\ref{fig:chisq}.  \rev{For $m_Q=500$ GeV, $\Lambda = 500$ eV}, this requirement is 94\% efficient for quirks and removes all SM tracks in our samples. \rev{For other $m_Q,\Lambda$ comparable efficiencies of $\approx80\%$ are found, except for the heaviest case $m_Q=5000$ GeV, due to their larger opening angles (however, 5 TeV quirks have a small cross section and are not produced at any appreciable rate).} To ensure a negligible background, these requirements can be tuned in data, but these studies suggest that such suppression is possible.  Further suppression is available by fitting two quirk candidate tracks simultaneously, if needed.  Additional discrimination power may be available from the momentum of the recoiling jet, which is correlated with the quirk pair momentum. 

To estimate the potential of discovering the quirk signal among the SM background, we estimate the number of reconstructed quirk events, $N_{\rm Reconstructed}$, as follows. Following Ref.~\cite{Knapen:2017kly}, we select quirk events that include a $p_\textrm{T}>200$ GeV jet to pass the High Level Trigger, and calculate the number of reconstructed quirk events as
\begin{equation}
    N_{\rm Reconstructed} = {\cal L}\times \sigma(Q\bar{Q}j,p_{\textrm{T}}^{\textrm{jet}}>200~{\rm GeV})\times \epsilon_{\rm total},
\end{equation}
where $\epsilon_{\rm total}$ is the total efficiency for reconstruction\rev{, selection and also includes the fraction of well-behaved quirks at this parameter point}. 
The efficiency $\epsilon_{\rm total}$  varies from $\lesssim$1\% to $\approx$ \rev{30}\% across the plane \rev{as shown in the left panel of} Fig.~\ref{fig:quirkeff}.  The right panel of Fig~\ref{fig:quirkeff} shows the expected quirk event rate $N_{\rm Reconstructed}$ in 300 fb$^{-1}$ of $pp$ collisions at the LHC, showing $N>3$ across $\Lambda\lesssim 5000$ eV for $m_Q\lesssim 700$ GeV. }

\section{Discussion}
\label{Sec:Discussion}

Reconstruction of quirk tracks in the region where the oscillation length ($d$) is larger than detector resolution, but smaller than the detector is expected to be very challenging. The application of ML-based tracking demonstrates a new-found capability to reconstruct such tracks with reasonable efficiency, in excess of 50\% across a range of parameter values where $d \approx 1-10$ cm.  For comparison, previous work which searches for co-planar hits to avoid fitting quirk trajectories~\cite{Knapen:2017kly} has a similar reconstruction efficiency for small oscillation lengths, $d < 1$ cm, but drops very quickly for cm scale oscillations, reaching $\approx 10\%$ for $m_Q,\Lambda$ = 800 GeV, 1000 eV. Training ML trackers explicitly on quirks can significantly extend the sensitivity range, complementing the co-planar approach. \rev{The technique presented in Sec.~\ref{Sec:Performance} is found to be sensitive to $m_Q\lesssim700$ GeV, for $100{~\rm eV}<\Lambda <5000{~\rm eV}$.}

One approach for the implementation of an ML-based quirk-finding algorithm is to analyze a set of events which are selected for other reasons, such as large \met.  But a more ambitious and much more challenging direction would be to generalize the core track-finding in the reconstruction pipeline so that it is capable of identifying helical and non-helical tracks, whether those are quirks or other unexpected particles.

Ref.~\cite{Knapen:2017kly} estimated the trigger efficiency for quirks directly to be low with existing tools, relying on primarily muon triggers. However, an ML-based tracker for quirks opens a new possibility of identifying quirks directly in the trigger, using knowledge distillation~\rev{~\cite{hinton2015distillingknowledgeneuralnetwork}} and hardware accelerators such as GPUs to fit the GNN within the trigger budget.

\section{Conclusions}

The analysis of data from particle physics experiments necessarily sacrifices some general sensitivity for the sake of practical computability, leaving us blind to many potentially striking signatures of new particles. An example of this is quirks, confining particles where the quirk mass, $m_Q$, is greater than the confinement scale, $\Lambda$, opposite to that of quarks. This hierarchy leads to an oscillating pair of non-helical tracks from the particles once they are pair produced at the LHC and propagate through the detector. Current tracking algorithms fail to reconstruct the quirk tracks when their oscillation is on the scale of the detector, ${\cal O} ({\rm cm})$. Larger oscillations would appear as two helical tracks, and shorter ones would appear as a nearly straight track through the detector.

Recent advances in ML-based reconstruction allow for new inroads into the previously inaccessible territory. In this work we have shown that graph neural network tracking can be trained to reconstruct quirk tracks with macroscopic oscillation lengths in the problematic centimeter range. We have generated quirk pseudodata across the $m_Q,\Lambda$ parameter space in a simplified model of the ATLAS detector and trained the {\sc Exa.TrkX} machine learning algorithm to reconstruct quirks. When trained on only SM tracks, the algorithm naturally fails to reconstruct tracks, but when instead trained on quirks the algorithm reaches comparatively large reconstruction efficiencies, $\sim 40-90\%$. We have also tested the robustness of the algorithm, by including background tracks from the associated jet in the signal process and found that these additional backgrounds do not significantly degrade our reconstruction performance \rev{or the sensitivity to discovering quirks. Effects from pile-up remain to be studied rigorously in future work, but existing implementations of GNN reconstruction in high pile-up (HL-LHC-like) conditions are performant with comparable sensitivity to traditional techniques \cite{ExaTrkX:2021abe, caillou2022atlas}. Although quirk trajectories are clearly more topologically complex than an SM hadron, we have no reason to suspect that their reconstruction is particularly sensitive to pile-up and noise effects, and remain optimistic that future studies in this direction will maintain the high efficiency demonstrated in this work.} The agnostic nature of the ML pipeline to the specific features of these tracks suggests a potentially broader capability to reconstruct non-helical tracks in a model-independent way. 

The application of such techniques to the existing vast LHC dataset might identify quirk candidates, which could be nearly background-free, as the probability for a SM track to exhibit such oscillations or of random hits creating such a path are near zero.

\section{Acknowledgements}

DW is funded by the DOE Office of Science. Y.Fang and Q.Sha are supported partially by Institute of High Energy Physics, Chinese Academy of Sciences under the innovative project on sciences and technologies with no.E3545BU210. DM is supported by the Danish Data Science Academy, which is funded by the Novo Nordisk Foundation (NNF21SA0069429). The work of MF~was supported in part by U.S.~National Science Foundation Grants PHY-2111427 and PHY-2210283, as well as by NSF Graduate Research Fellowship Award No. DGE-1839285.
The authors are grateful to Simon Knapen and Michele Papucci for useful discussions and for sharing the quirk trajectory code from ~\cite{Knapen:2017kly}. We are also grateful to Markus Elsing and Xiangyang Ju for their comments and expertise.
\appendix
\section{ SM efficiency}
\label{ref:app_smeff}

\begin{table}[h]
    \centering
     \caption{Efficiency to reconstruct quirk tracks for various values of quirk mass ($m_Q$) and confinement scale ($\Lambda$), when training on the SM trackers and inference on the quirk trackers based on 25-layer and {\it well-behaved} setting.}
    \label{tab:eff_lambda_Mq_forSM}
    \begin{tabular}{c|cccc}
    \hline \hline
    & \multicolumn{4}{c}{$m_Q$ (GeV)} \\
    $\Lambda$ (eV) & 100 & 500 & 1000 & 5000 \\
    \hline
    100   & 81.3\% & 89.9\% & 93.2\% & 80.5\% \\
    500   & 19.0\% & 7.0\%  & 5.4\%  & 2.2\%  \\
    1000  & 18.8\% & 12.8\% & 4.8\%  & 13.0\% \\
    2000  & 15.4\% & 11.8\% & 31.2\% & 8.0\%  \\
    3000  & 18.6\% & 32.2\% & 24.6\% & 15.2\% \\
    4000  & 17.0\% & 21.0\% & 29.6\% & 18.8\% \\
    5000  & 29.8\% & 24.6\% & -      & -      \\
    \hline \hline
    \end{tabular}
\end{table}

\begin{table}[]
   \centering
    \caption{Score cut for different points}
   \begin{tabular}{cccccc|c}
   \hline \hline
        $m_Q$ (GeV)  & $\Lambda$(eV) &$\bar{\gamma}_{\rm }$ & $\sigma_\gamma$ & $d$ [cm] &  Efficiency  &Score cut\\
        \hline
         100 & 100 & 4.4 & 3.4& 670 & 91.0\%    & 0.9\\
          & 500  & 3.7 & 2.7 & 21.6 & 82.8\%    & 0.9\\
          & 1000&  3.1& 2.05 & 4.2  & 77.8\%    & 0.9\\
          & 2000& 3.0 & 2.0 & 1.0   & 60.4\%    & 0.6\\
          & 3000& 3.0 & 1.9 & 0.44  & 35.6\%    & 0.5\\
          & 4000& 2.9 & 1.8 & 0.24  & 34.5\%    & 0.5\\
          & 5000& 2.9 & 1.7 & 0.15  & 24.2\%    & 0.5\\
         \hline
         500 & 100& 1.9 & 0.7 & 896 & 92.0\%    & 0.9\\ 
          & 500&1.8 &0.6 & 31       & 79.0\%    & 0.9\\
          & 1000&1.7 &0.6 & 7.3     & 64.3\%    & 0.9\\
          & 2000&1.7 & 0.5& 1.7     & 60.9\%    & 0.6\\
          & 3000&1.7 &0.5 & 0.8     & 59.6\%    & 0.6\\ 
          & 4000&1.6 &0.5 & 0.4     & 42.6\%    & 0.5\\
          & 5000&1.6 &0.5 & 0.2     & 39.2\%    & 0.5\\
                  \hline
         1000 & 100&1.5 &0.3 & 950  & 94.6\%    & 0.9\\
          & 500& 1.4& 0.3& 32       & 63.6\%    & 0.9\\    
          & 1000&1.4 &0.3 & 7.6     & 62.7\%    & 0.6\\
          & 2000&1.4 &0.3 & 1.8     & 69.2\%    & 0.7\\
          & 3000& 1.3&0.2 & 0.8     & 54.1\%    & 0.6\\
          & 4000& 1.3&0.2 & 0.4     & 59.7\%    & 0.6\\ 
                  \hline
         5000 & 100&1.04 &0.03 & 420    & 84.8\%    & 0.9\\
          & 500&1.03 &0.02 & 11         & 69.8\%    & 0.9\\     
          & 1000& 1.03& 0.02& 2.7       & 69.5\%    & 0.6\\  
          & 2000& 1.03 & 0.02 & 0.7     & 49.2\%    & 0.5\\
          & 3000& 1.03& 0.02& 0.3       & 36.4\%    & 0.5\\
          & 4000&1.03 & 0.02& 0.2       & 34.6\%    & 0.3\\
         \hline
         \hline
   \end{tabular}
   \end{table}

\clearpage
\bibliography{quirks}
\end{document}